\documentclass[
 reprint,
 amsmath,amssymb,
 aps, pre
]{revtex4-2}

\usepackage{graphicx}
\usepackage{dcolumn}
\usepackage[caption=false]{subfig}
\usepackage{bm}
\usepackage{float}
\usepackage{mathtools,amsmath,amssymb}
\usepackage{hyperref}

\preprint{}

\begin{document}

\title{
Paradise-disorder transition in structural balance dynamics on Erd\"os-R\'enyi graphs}
\author{Krishnadas Mohandas}
\email{krishnadas.mohandas.dokt@pw.edu.pl}
 \author{Krzysztof Suchecki}
 \email{krzysztof.suchecki@pw.edu.pl}
\author{Janusz A. Ho{\l}yst}
 \email{janusz.holyst@pw.edu.pl}
 \affiliation{
Faculty of Physics, Warsaw University of Technology\\
Koszykowa 75, PL-00-662 Warsaw, Poland
}

\begin{abstract}
    Structural balance has been posited as one of the factors influencing how friendly and hostile relations of social actors evolve over time.
    This study investigates the behavior of the Heider balance model in Erd\"os-R\'enyi random graphs in the presence of a noisy environment, particularly the transition from an initially entirely positively polarized paradise state to a disordered phase.
    We examine both single-layer and bilayer network configurations and provide a mean-field solution for the average link polarization that predicts a first-order transition where the critical temperature scales with the connection probability $p$ as $p^2$ for a monolayer system and in a more complex way for a bilayer.
    We show that to mimic the dynamics observed in complete graphs, the intralayer Heider interaction strengths should be scaled as $p^{-2}$, while the interlayer interaction strengths should be scaled as $p^{-1}$ for random graphs.
    Numerical simulations have been performed, and their results confirm our analytical predictions, provided that graphs are dense enough.
\end{abstract}
\maketitle

\section{Introduction}
\label{sec:introduction}
The Heider Structural Balance Theory (SBT) posits that relations between humans tend towards so-called balanced states \cite{heider1946attitudes}, which means states where a friend of a friend is also a friend, a friend of an enemy is an enemy, an enemy of a friend is an enemy and enemy of an enemy is a friend.
This intuitive set of rules stems from observations of relations in social systems and implies the existence of triadic interactions.
When interactions exist between every pair of people, it results in two possible fully balanced states: a paradise where everybody is a friend of everybody, or a split into cliques of friends that are hostile towards each other \cite{cartwright1956structural,marvel2009landscape,marvel2011continuous}.
Except for specific circumstances, the most likely state to emerge is the split clique state, making SBT a simple explanation of the propensity of humans to divide into opposed groups.\\

Although the theory stipulates what kinds of triads are balanced, it does not define specific rules for the dynamics of the relations.
There have been investigations of the structural balance in a model where imbalanced triads are changed to balanced ones according to set probabilities \cite{antal2005dynamics,antal2006social,abell2009structural}, as well as more recent studies where the tensions in the system due to imbalanced relationship triads are assigned energy and the dynamics is assumed to follow statistical behavior of systems under thermal noise \cite{belaza2017statphys, rabbani2019mean, malarz2022mean}.
While this is also an assumption, such an approach has several advantages.
The specific nature of thermal noise dynamics in a system with defined energy is, in general, well-known in physics.
The approach does not introduce any biases in sampling system phase space due to particular dynamical rules and usually features an equilibrium state.\\

Various models based on SBT principles have been developed to explain dynamics within signed networks \cite{summers2013active,facchetti2011computing,doreian2015structural,yang2022promotive}. 
While the structure of ties in real social systems is far from simple, most investigations of the structural balance assume relations exist between every pair of agents \cite{burghardt2020dyadic,pham2022empirical,gorski2020homophily}, which means that the structure of interactions is a simple complete graph.
This may be due to the fact that triadic interactions require either a high density of connections or a complicated topology with a high local clustering coefficient, as typical sparse random graphs are almost devoid of triangles required for Heider interactions to occur.
Erd\"os-R\'enyi random graphs, where every pair of nodes is connected with fixed probability $p$ \cite{erdos1959random}, are well researched, their intrinsic topological properties including motifs \cite{winkler2013motifs,itzkovitz2005subgraphs,shi2014identifying} and percolation properties \cite{callaway2000robustness}, as well as behavior of various models \cite{dorogovtsev2002ising,dembo2010ising,vilone2004voter,pereira2005majority,pournaki_order-disorder_2023} are well known.
Typically, the models with two-body interactions like the Ising model will show scaling of critical properties with a mean degree, which means it will scale with random graph density $p$ linearly \cite{dembo2010ising, dorogovtsev2002ising}.
There are also models, like the voter model, that lack any critical properties related to mean degree, and their behavior is not significantly affected by parameter $p$.
In the voter model's case, it is the apparent dimensionality of the network \cite{frachebourg1996exact} that plays the main role, although the graph density influences the expected time to reach an absorbing state in finite systems \cite{sood2005voter}.
On the other hand, the impact of sparse topology on the dynamics of three-body Heider interactions is currently not well explored.
Two notable works on that topic have been recently done by Masoumiet al. \cite{masoumi2022modified}, and Malarz et al. \cite{malarz2023heiderer} that investigate the behavior of the SBT model in random graphs.
The first presents an analytical approach and compares it with numerical simulations focusing mainly on the temperature of transition from disorder to order.
The second work also regards the same transition and contains the results of extensive numerical simulations but provides no analytical solutions.\\

In this paper, we investigate the behavior of SBT on Erd\"os-R\'enyi random graphs, with the aim to analytically describe a transition from an ordered paradise state to a disordered state and to understand how the sparsity of the graph affects this transition.
While \cite{masoumi2022modified,malarz2023heiderer} only considers the lowest temperature where disorder can exist, our work focuses on the other, highest temperature where the order can be preserved.
Both temperatures are not the same due to system multistability.
We use the mean-field approach to derive an analytical approximation for the critical temperature for the existence of an ordered state in the model and use numerical Monte Carlo simulations to verify it.
We also investigate the behavior of the model in a system composed of two interacting layers \cite{mohandas2024critical, gorski_destructive_2017}, each being a random graph.
We find the critical temperature to depend on the network density as $p^2$ for a single network and in a more complex fashion for the two-layer case.
In particular, we note that the system behaves similarly to the complete graph, except with Heider interaction strength scaled as $p^{-2}$ and interlayer coupling strength scaled as $p^{-1}$ for the bilayer network case.
In comparison with previous works regarding SBT on random graphs, \cite{masoumi2022modified, malarz2023heiderer}, our paper provides explicit mean-field predictions for the critical temperature of order-disorder transition and an analytical approach for duplex random graphs.

\section{Model}
\label{sec:model}
\begin{figure}
    \centering
    \includegraphics[width=0.9\linewidth]{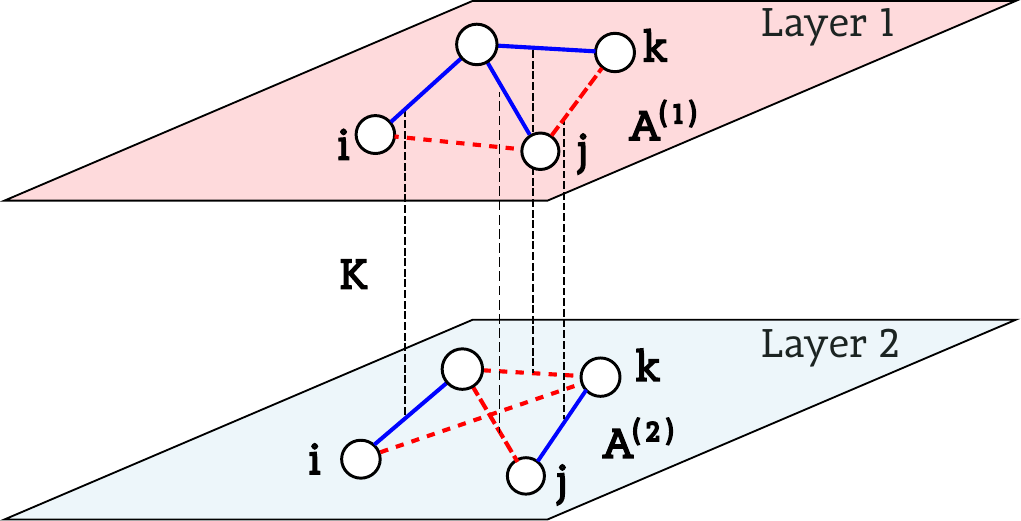}
    \caption{An illustration of interactions between signed links of a bilayer ER network. Positive links are represented by solid blue lines, and negative links are represented by dashed red lines. The structural balance drives the network to a balanced state, which is governed by the intralayer interactions as well as the interlayer interactions. When the interlayer coupling strength $K=0$, the system is reduced to two independent single-layer ER networks.}
    \label{fig:bilayer-illustration}
\end{figure}

Consider an  Erd\"os-R\'enyi (ER) network with $N$ agents and connection probability $p$ connecting any two nodes $i$ and $j$ with a link $ij$.
Implementing structural balance on the ER network requires the links to have signs that correspond to polarizations of social relations $x_{ij}$ between two agents.  
For simplicity, we will assume that for connected nodes $x_{ij} = \pm 1$, and if there are no links connecting the nodes, then $x_{ij}=0$.
Note that link states $x_{ij}=0$ are a way to represent missing links, and they do not represent true link states, and the network topology is quenched and does not change in time.\\
Since the structural balance theory considers the stable triads to be balanced, the energy associated with such a balanced triad is $-A$.
The parameter $A > 0$ represents the Heider interaction strength between the links in the network.\\
As in \cite{malarz2022mean}, the Hamiltonian $H$ of the system is
\begin{equation}\label{eq:hamiltonian1}
    H = -A\sum_{i<j<k}^N x_{ij}x_{jk}x_{ki} 
\end{equation}
where $N$ is the size of the network.\\
Degenerate ground states for this Hamiltonian correspond to cases when every triad of connected nodes $(ijk)$ is {\it balanced} \cite{heider1946attitudes,cartwright1956structural,marvel2009landscape,marvel2011continuous,antal2005dynamics,antal2006social}, i.e., either all links    $(ij)$, $(jk)$, $(ki)$ are positively polarized, or there are two negative links and one positive, e.g. $x_{ij}=x_{jk}=-1$ and  $x_{ki}=1$.
In both cases, the product of all link polarizations in the triad is positive $x_{ij}x_{jk}x_{ki}=1$, and the energy of such a triad equals $-A$.
Triads consisting of one negative link and two positive links or of three negative links are unbalanced, with products of all their link polarizations negative $x_{ij}x_{jk}x_{ki}=-1$.
Thus, their energies are equal to $A$ and can be considered as excited states \cite{rabbani2019mean,malarz2022mean} in statistical ensembles. 
Since we consider an ER graph, some links $x_{ij}$ are missing, and the energy of an associated triad contributing to the Hamiltonian is zero.
For a bilayer ER network, an assumption is made that nodes are identical in each layer and coupling exists only between link states of the same node pairs in different layers (see Fig. \ref{fig:bilayer-illustration}).
Considering the topology of a random network, the link connecting a node pair $ij$ in the first layer might be absent in the second layer. Then, the interlayer interaction between the node pairs would not exist.
The tension arising from discrepancies between relations in different layers for the same pair of agents can also be represented as energy.
Specifically, if the relations are the same, the energy is reduced by a positive amount $K$, whereas if the relations are different, the energy increases by $K$.\\
The Hamiltonian for a bilayer network, with links $x_{ij}^{(\alpha)}$ and interaction strengths $A^{(\alpha)}$ in layers $\alpha=1,2$, is
\begin{equation}
\begin{aligned}
    H =& -A^{(1)}\sum_{i<j<k}^N x_{ij}^{(1)} x_{jk}^{(1)} x_{ki}^{(1)}-A^{(2)}\sum_{i<j<k}^N x_{ij}^{(2)} x_{jk}^{(2)} x_{ki}^{(2)}\\
    &- K\sum_{i<j}^N x_{ij}^{(1)} x_{ij}^{(2)} \label{eq:hamiltionian2}
\end{aligned}
\end{equation}
Ground states of the Hamiltonian \ref{eq:hamiltionian2} correspond to cases when all existing triads $(ijk)$ are balanced in both layers, i.e. $x_{ij}^{(\alpha)} x_{jk}^{(\alpha)} x_{ki}^{(\alpha)}=1$ and link polarizations are the same in both layers $x_{ij}^{(1)} =x_{ij}^{(2)}$.
Note that there are no internal node attributes, and the entire system dynamics considers only changes in link signs $x_{ij}^{(\alpha)}$.
The state of the system will change according to the Hamiltonian in the presence of temperature $T$, which introduces random noise to the dynamics.
The temperature represents the unpredictability of human behavior or tolerance towards imbalanced relations in the social systems modeled.
The model has a special ground state where all existing links $x_{ij}^{(\alpha)}$ are positive, which is called the {\it paradise} state.
Other ground states also exist \cite{antal2005dynamics,antal2006social,malarz2022mean}, with internally positively connected groups and hostile relations between groups, but we do not investigate them in this paper.
Due to thermal fluctuations, at any temperature $T > 0$, the system settles into an equilibrium different than the ground state.

\section{Analytical approach}
\label{sec:analytical}
\subsection{Mean-field approximation for single ER network}\label{sec:meanfield single}
Let us start with a monolayer ER graph of size $N$ and with the connection probability $p$, where the Hamiltonian is described by Eq.(\ref{eq:hamiltonian1}).
The problem of mean-field description of the Hamiltonian-driven structural balance for a complete graph has been solved before \cite{malarz2022mean}.
Our situation differs since a fraction of $1-p$ of the links between agents does not exist.
Since the ER network is uncorrelated and features fewer links and triads than the complete graph for $p<1$, one can expect that the model will behave similarly but with a lower effective interaction strength and, in effect, a lower critical temperature.
Let us assume that the nodes $i$ and $j$ are connected by a link.
The polarization $x_{ij}$ of this link is a stochastically fluctuating variable.
Since we consider the system in a constant temperature $T$, the probability $P(x_{ij})$ of finding the system in a microstate $x_{ij}=\pm 1$ is described by Boltzmann distribution. 
\begin{equation}
    P(x_{ij})= \frac{\exp[- h(x_{ij})/T]}{\sum_{x_{ij}=\pm 1}\exp[- h(x_{ij})/T]} \label{eq:canonical}
\end{equation}
where,
\begin{equation}
    h(x_{ij})= -A x_{ij} \sum_{k\neq i,j}^N x_{jk}x_{ik}
\end{equation}
is the Hamiltonian part dependent on the state of the link $ij$.
The expected value $\langle x_{ij} \rangle$ can be expressed as
\begin{equation}\label{eq:meanx}
    \left\langle x_{ij} \right\rangle =\sum_{x_{ij}= \pm 1} P(x_{ij})x_{ij}
\end{equation}
Following the mean-field approach developed in \cite{malarz2022mean}, the link $x_{ij}$ interacts with the mean-field that is proportional to $x^2$  rather than with the product of specific link polarizations  $x_{jk}$ and $x_{ki}$.
The value of this mean field reflects the mean state of links across the entire system, including the link $ij$ itself, as we consider them all statistically the same, meaning $\langle x_{ij} \rangle \equiv x$.
Since we want to consider the polarization $\langle x_{ij} \rangle$ of existing links, this average will include only links with states $x_{ij}=\pm 1$.
Nonexistent links that we formally considered to have $x_{ij}=0$ are not included in the average $x$, so the mean link value for the entire network can be written as
\begin{equation}
    x =  \sum_{i<j}^N \left\langle x_{ij} \right\rangle/ \sum_{i<j}^N \left\langle|x_{ij}|\right\rangle
\end{equation}
where the denominator is an expression equal to the number of links $x_{ij}=\pm 1$, meaning all existing links.
We refer to this mean link sign value $x$ as the polarization of the network.
Since the sum in Hamiltonian $h(x_{ij})$ is over all node pairs $jk$ and $ki$, and the mean-field $x^2$ does not include nonexistent links, we must account for that.
Each link $jk$ and $ki$ exists with a probability $p$, which means that for the product $x_{jk}x_{ki}$ to be non-zero, both links must exist, which happens with a probability $p^2$ since the existence of links is uncorrelated in ER graph.
Consequently, we average over $1-p^2$ products with zero value and $p^2$ products with non-zero value and can therefore express the energy of a given state $x_{ij}$ as:
\begin{equation}\label{eq:energy}
    h(x_{ij}) \stackrel{\mathrm{MF}}{=} - A x_{ij} (N-2) p^2 x^2 
\end{equation}
This is the same equation as the mean-field value for a complete graph derived in \cite{malarz2022mean} except multiplied by a constant factor $p^2$, allowing us to follow the same reasoning further.
From Eq. (\ref{eq:canonical}), (\ref{eq:meanx}) and (\ref{eq:energy}), a self-consistent equation for the mean link polarization $x$ as the function of the temperature $T$ can be obtained,
\begin{equation}\label{eq:mfequation}
    x =  \tanh[{a(T)x^2}]
\end{equation}
where $a =  \frac{Ap^2M}{T}$, and $M=(N-2)$ is the number of triads containing a given link in a complete graph.
This equation, along with an equation for the tangency of the right-hand side of (\ref{eq:mfequation}) to the diagonal
\begin{equation}
    \frac{d}{dx} \tanh({a_cx_c^2}) =1\label{eq:mftangequation}
\end{equation}
form a pair of algebraic equations for $x_c$ and $a_c$ when the discontinuous transition from an ordered to a disordered state occurs \cite{malarz2022mean}.
Unfortunately, they have no closed-form analytical solution due to their transcendental nature, but we can see that the only difference from the case of a complete graph studied in \cite{malarz2022mean} is the re-scaling of interaction strength $a$ by $p^2$.\\

Figure \ref{fig:subfig4a} shows the numerical solution of mean polarization for the single layer ER network obtained from the mean-field solution (Eqs. \ref{eq:mfequation}, \ref{eq:mftangequation}) for different values of connection probability $p$.
We will assume that the initial condition of the system is a paradise state.
In such case, one can anticipate that as the temperature $T$ increases, the mean link polarization in equilibrium decreases from a fully polarized state ($x=1$) and reaches $x_c$ where a discontinuous transition occurs at $T_c$, beyond which the mean polarization becomes zero.
The numerical solution \cite{malarz2022mean} yields $x_c \approx 0.79638$, independent of model parameters.
The critical temperature $T_c$ can be expressed as
\begin{equation}\label{eq:criticalT1}
    T_c = \left(\frac{1}{a_c}\right)Ap^2(N-2)
\end{equation}
where $a_c \approx 1.71649$ is also independent of model parameters.
From the above equation, a normalized critical temperature $\Tilde{T}_c = T_c/AM$ can be written as 

\begin{equation}
\Tilde{T}_c=(1/a_c)p^2\approx 0.58258 p^2.
\end{equation}
\subsection{Mean-field approximation for bilayer ER network}
\label{sec:meanfield-bilayer}
\begin{figure}
   \includegraphics[width=\columnwidth]{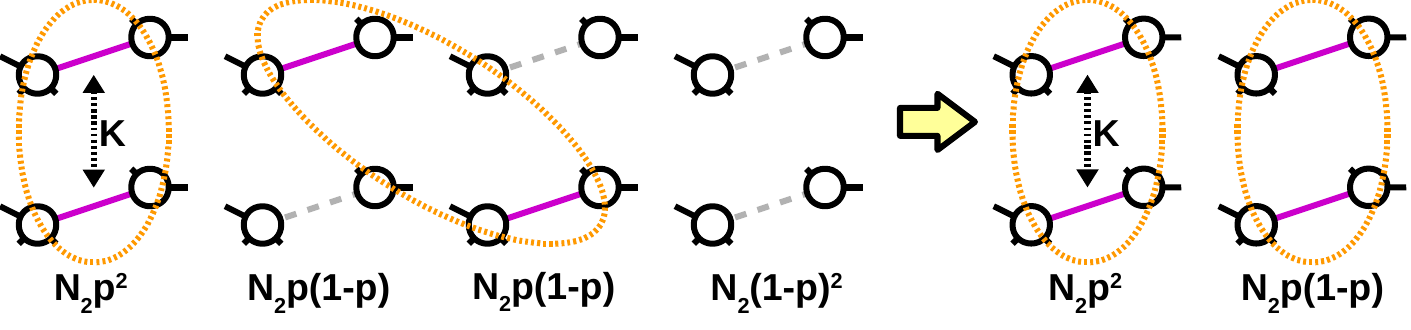}
   \caption{ In bilayer Erd\"os-R\'enyi graph with $N$ nodes and probability $p$ for each link to exist, there are $N_2 p^2$ pairs of nodes that have links in both layers, $2 N_2 p (1-p)$ pairs with link in only one layer and $N_2 (1-p)^2$ node pairs without any links, where $N_2=N(N-1)/2$ is the total number of node pairs. We can pair up (encircled in orange) links into link pairs, resulting in $N_2 p^2$ coupled node pairs that remain coupled with interaction strength $K$ and $N_2 p(1-p)$ pairs that are not coupled due to being assembled out of links that connect in fact different node pairs and do not interact. Since those pairs do not interact and we consider mean-field approximation only, the fact the links of those pairs belong to different node pairs does not matter. This mixture of interacting and non-interacting link pairs can be used for the mean-field description of the model in bilayer random graphs.}
   \label{fig:explainbilayermf}
\end{figure}

The extension of mean-field approach to a bilayer network composed of two complete graph layers is discussed in \cite{mohandas2024critical}, where a pair of coupled links $\vec{x}_{ij}=[x^{(1)}_{ij}, x^{(2)}_{ij}]$ is considered as the elementary subsystem interacting with the mean-field, instead of a single link $x_{ij}$.
Applying the same methodology for a bilayer random graph is not trivial because not all node pairs $ij$ have links in both layers.
If we consider two independent random graphs with a probability $p$ of each potential link existing and treat them as two layers of the same duplex graph, then we have $\tfrac{N(N-1)}{2}p^2$ node pairs 
that have links in both layers, $\tfrac{N(N-1)}{2}2p(1-p)$ node pairs featuring a link only in one of the layers, and $\tfrac{N(N-1)}{2}(1-p)^2$ node pairs without any links (see Fig. \ref{fig:explainbilayermf}).
Since we want to use the mean-field approximation, we can forget about specific positions of links in the structure and pair up all single links in layers $1$ and $2$ into effectively non-interacting link pairs.
We end up with $\tfrac{N(N-1)}{2}p^2$ link pairs coupled with constant $K$ and $\tfrac{N(N-1)}{2}p(1-p)$ non-interacting link pairs where effective $K=0$.
This means there is a fraction $p$ of all link pairs that have interlayer interactions and $1-p$ fraction that do not.
This approximation works only if both layers share the same density of links $p$ and, therefore, have the same mean number of links.\\
Now, we can follow the same methodology as for the bilayer complete graph \cite{mohandas2024critical}, except with only a fraction $p$ of link pairs featuring interlayer interactions, which means that the mean interaction strength between layers in $K p$ instead of $K$.
Assuming symmetry between layers $A^{(1)}=A^{(2)}=A$ and mean link pair polarization across the system
\begin{equation}
    \vec{x} \equiv [x^{(1)}, x^{(2)}] =  \left[ \frac{\sum_{i,j}^N x_{ij}^{(1)}}{\sum_{i,j}^N |x_{ij}^{(1)}|},  \frac{\sum_{i,j}^N x_{ij}^{(2)}}{\sum_{i,j}^N |x_{ij}^{(2)}|}\right]
\end{equation}
we can then approximate the energy tied to the state $\vec{x}_{ij}$ as
\begin{multline}\label{eq:energyxij}
    h(\vec{x}_{ij}) \stackrel{\mathrm{MF}}{=} -A \left( x_{ij}^{(1)} + x_{ij}^{(2)} \right) M p^2 x^2
    -K p x_{ij}^{(1)}x_{ij}^{(2)}  
\end{multline}
where $M=N-2$.\\
We can see that in the mean-field approach for bilayer ER graphs, the energy tied to a pair of links is the same as for bilayer complete graph \cite{mohandas2024critical}, except for two differences: the intralayer interaction strength is scaled by $p^2$, just as in the case of a single-layered network, and the interlayer interaction strength scaled by $p$ due to only fraction of link pairs interacting.
Since this is just a re-scaling of both effective interaction strengths, the behavior of the system should be the same as in the case of the complete graph, except with a modified critical temperature.\\
Following results of \cite{mohandas2024critical} the critical temperature is therefore
\begin{equation}\label{eq:CriticalTemp2L}
     T_c = Ap^2M\left(\frac{1}{\ln{z}}\right)\left( \frac{D^{2}(z^{4}-1)}{D^{2}(z^{4}+1)+2z^{2}} \right)^2
\end{equation}
where $D=\exp[{K/(ApM)}]$ and $z=z(D)$ is a solution of a transcendental equation
\begin{equation} \label{eq:transcendentalEqn}
    8 \ln z = \frac{(z^{4}-1)(z^{4}D^{2}+2z^{2}+D^{2})}{z^{2}(z^{4}+2z^{2}D^{2}+1)}
\end{equation}
that has the same form as the complete graph, except $D$, which depends on $p$.
The normalized critical temperature $(\Tilde{T}_c = T_c/AM)$ thus depends on the value of $p$.
Initially, the relationship between $\Tilde{T}_c$ and $p$ may appear quadratic, suggesting that the critical temperature rises with the square of  $p$.
However, considering the dependency of $D$ on $p$, the relationship is more complex.
For $D=1$ (corresponding to $K=0$), the $\Tilde{T}_c$ becomes equivalent to non-interacting layers and features simple quadratic dependence on $p$.\\

To conclude the results for the mean-field description, if the intralayer interaction strength $A$ is scaled to $A/p^2$ and the interlayer interaction strength is scaled from $K$ to $K/p$, then the system statistical properties are independent of $p$, including the normalized critical temperature $\Tilde{T}_c$ and are the same as the case of a complete graph ($p=1$).
This is shown in the inset of Fig. \ref{fig:criticaltemp2L}.

\section{Numerical simulation}
\label{sec:numerical}
Numerical Monte Carlo simulations of the model for a single layer and bilayer ER network have been carried out to verify the analytical calculations presented in the previous section.
We employ an asynchronous Metropolis algorithm, where $pN (N-1)/2 $ elementary updates are considered as a time step.\\
For a single-layered ER network, in each update, a random pair of connected agents $ij$ is chosen, the difference of energy $\Delta E$ between configuration with inverted $x_{ij}$ and current configuration is calculated, and the sign of the link is inverted with probability $\min(1,e^{-\Delta E/T})$.
For the bilayer network, in each update, two pairs of connected agents $ij$ and $kl$ are chosen randomly, each from a different layer.
The energy difference $\Delta E$ is calculated between a configuration with both link signs inverted and the current configuration, including both intralayer Heider interactions and interlayer coupling.
Finally, both link signs $x_{ij}^{(1)}$ and $x_{kl}^{(2)}$ are inverted with probability $\min(1,e^{-\Delta E/T})$.
The simulations were conducted across $50$ realizations of ER graph, ranging from a sparser network (with $p = 0.1$ ) to a complete graph $( p = 1)$, while systematically varying the temperature from $\Tilde{T} = 0$ in increments of $\Delta \Tilde{T}$.
This iterative process aims to determine the normalized critical temperature $(\Tilde{T}_c)$ at which the mean link polarization $x$ and normalized intralayer energy become zero.
The exact value of $\Tilde{T}_c$ is identified as the maximum of the slope obtained from the energy data averaged over $50$ realizations, wherein the system's energy changes from a ground state to a disordered state with higher energy (see Appendix \ref{sec:appendix}).
The standard error is also calculated from all the realizations.\\
To obtain the simulation results, the normalized intralayer and interlayer energies are calculated from the Hamiltonian (\ref{eq:hamiltionian2}).
The intralayer energy for each layer is normalized to represent the average energy per existing triad, given by 
\begin{equation}\label{eq:normalizedE}
    \Tilde{E}_{\alpha} = \frac{-A^{(\alpha)}\sum_{i<j<k}^N x_{ij}^{(\alpha)}x_{jk}^{(\alpha)}x_{ki}^{(\alpha)}}{ A^{(\alpha)}N_\Delta^{(\alpha)} }
\end{equation}
where $A^{(\alpha)}$  is the interaction strength for layer $\alpha$ and $N_\Delta^{(\alpha)}$ is the number of existing triads in layer $\alpha$ (mean value $\tfrac{N(N-1)(N-2)}{6}p^3$).
We fix the value of intralayer interaction strength $A^{(\alpha)}=1$, and the interlayer energy is scaled to the average energy value per interacting link pair given by,
\begin{equation}\label{eq:normalizedEK}
    \Tilde{E}_K = \frac{-K\sum_{i<j}^N x_{ij}^{(1)}x_{ij}^{(2)}}{K N_K }
\end{equation}
where $K$ is the interlayer interaction strength and $N_K$ is the number of existing coupled link pairs (mean value $N(N-1)p^2/2$.

\subsection{Critical behavior of a single layer ER network}
\label{sec:numerical-single}
\begin{figure*}
    \subfloat[\label{}]{
    \includegraphics[width=0.47\linewidth]{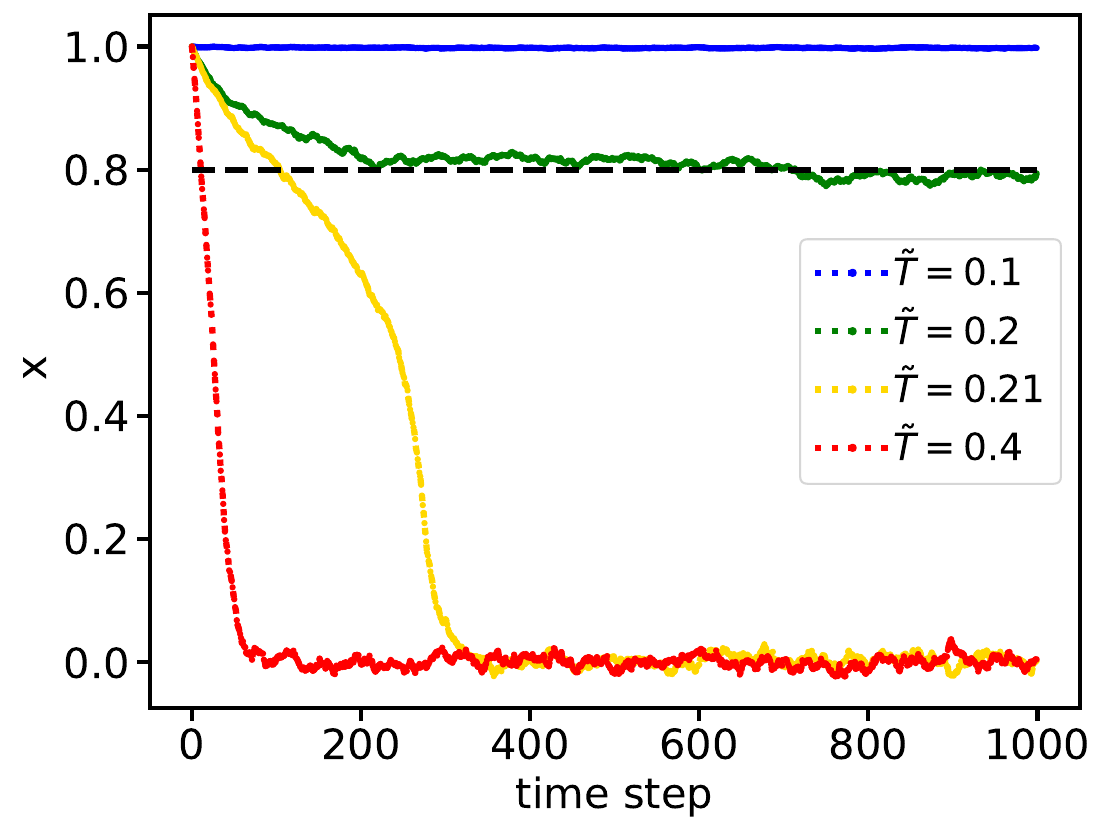}
    }\hfill
    \subfloat[\label{}]{
    \includegraphics[width=0.48\linewidth]{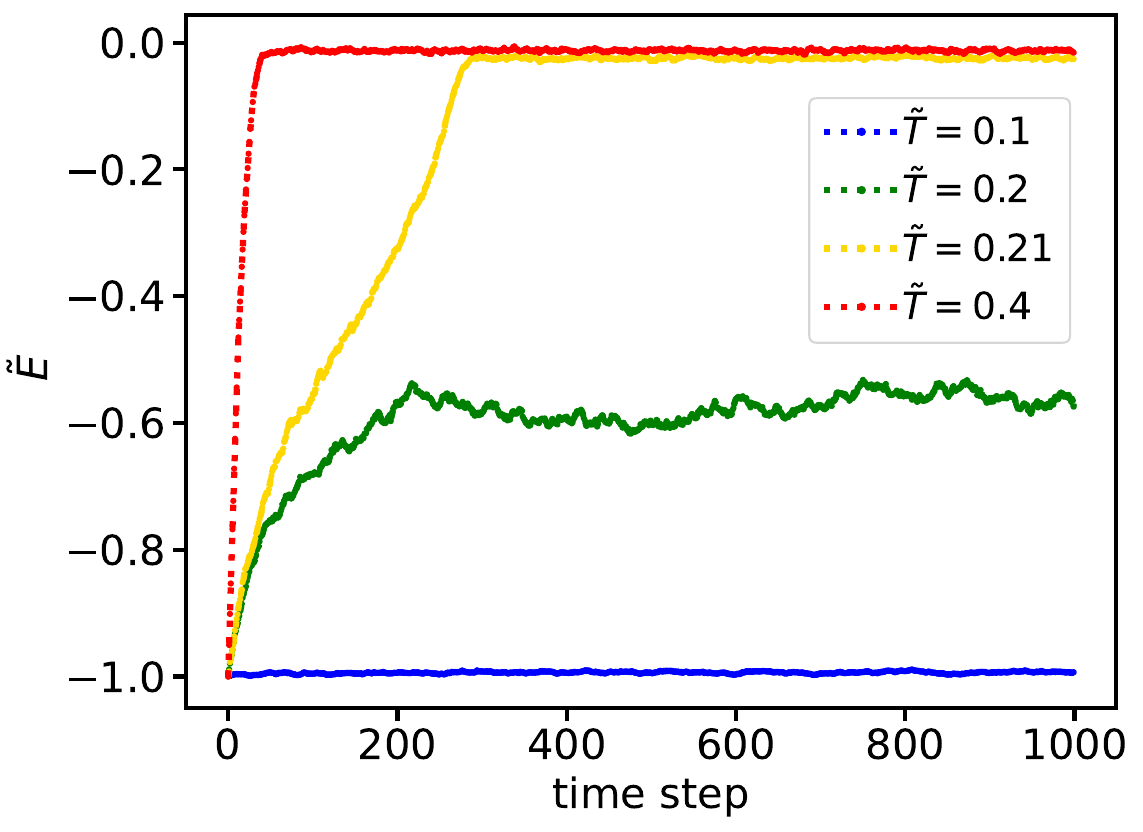}
    }
    \caption{Examples of the time evolution of the average values of all links’ polarization for temperatures below (blue and green lines) and above (yellow and red lines) the normalized critical temperature $(\Tilde{T}_c)$ for a network size $N=200$ with $p=0.6$.
    For $\Tilde{T}=0.2$, fluctuations are observed around values that are close but always larger than the critical solution $x_c$ (dashed black line), however for a slightly higher temperature, the asymptotic values of mean link polarization show a discontinuous transition.
    The starting point of all simulations is the paradise state.
    The normalized critical temperature $\Tilde{T}_c$ is found between the two temperature values $\Tilde{T}=0.2$ and $\Tilde{T}=0.21$.}
    \label{fig:time-evolution}
\end{figure*}
\begin{figure}
    \subfloat[\label{fig:subfig4a}]{
    \includegraphics[width=0.9\linewidth]{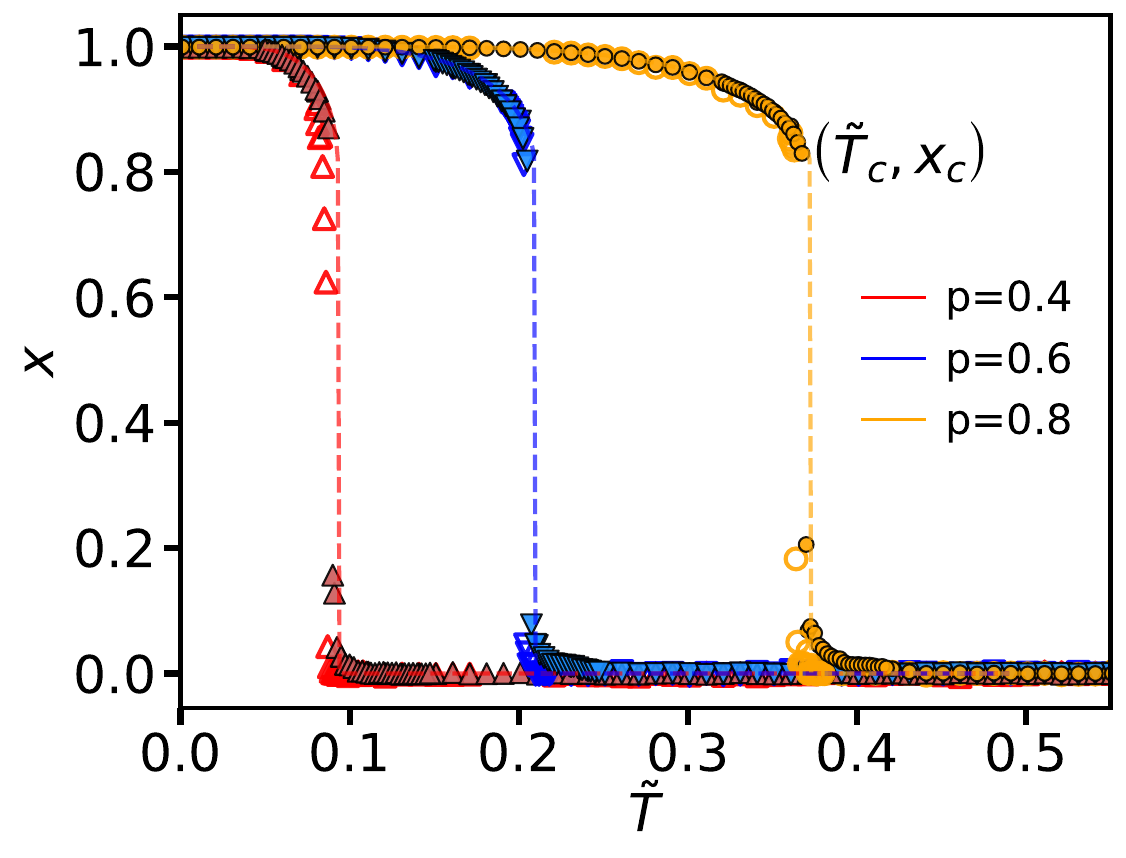}
    }\hfill
    \subfloat[\label{fig:subfig4b}]{
    \includegraphics[width=0.9\linewidth]{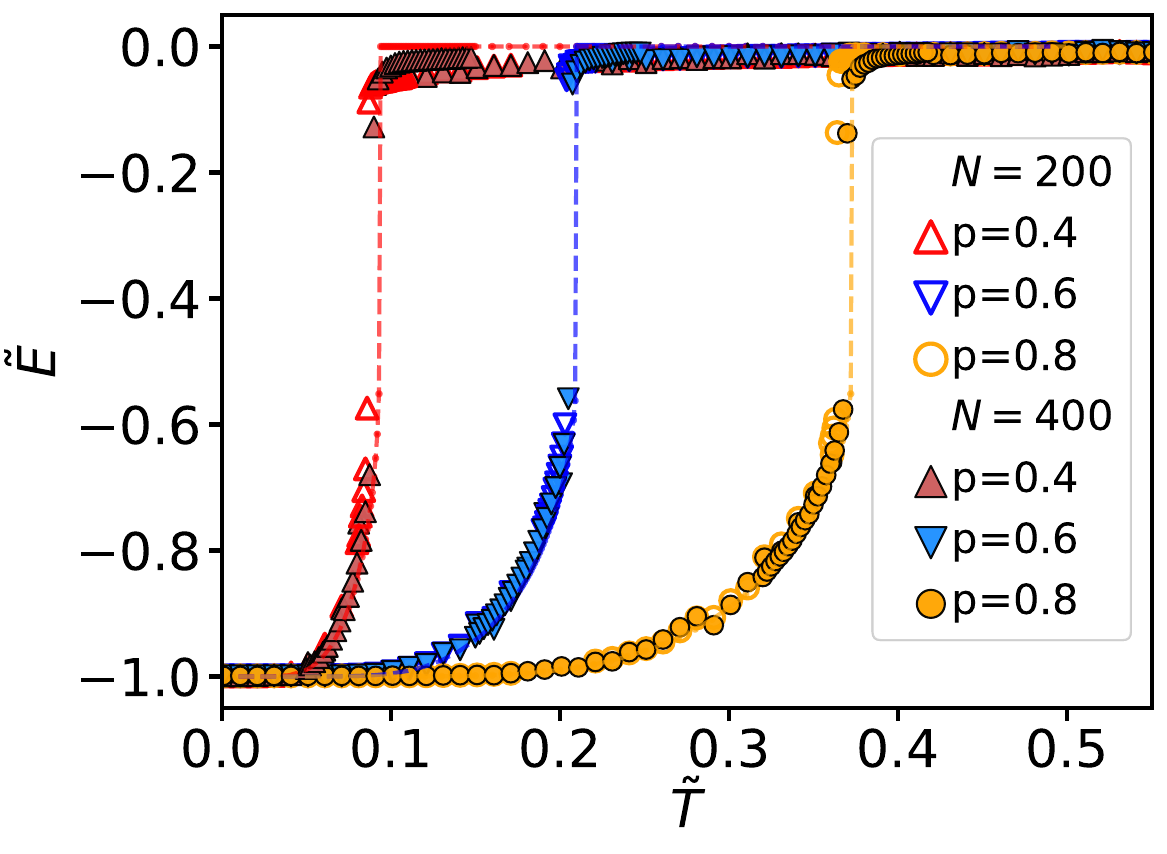}
    }
    \caption{Mean-field solution obtained analytically (lines) and the simulation results (triangles and circles) for a single layer ER network where a first-order transition is observed at the normalized critical temperature $\Tilde{T}_c$. The normalized critical temperature $\Tilde{T}_c$ decreases with the decreasing value of $p$ since the network is sparser and every link is influenced by a smaller number of triads. Panel (b) shows the corresponding normalized energy (Eq.\ref{eq:normalizedE}) of the simulated network. The results are displayed for $p=0.4, 0.6, 1$ and $N=200, 400$, where the transition point moves from left to right for increasing values of $p$. Both panels share the same legend. It is worth noting that normalized critical temperature values $\tilde{T}$ are independent of network size $N$.}
    \label{fig:meanfield}
\end{figure}
\begin{figure}
    \centering
    \includegraphics[width=1\linewidth]{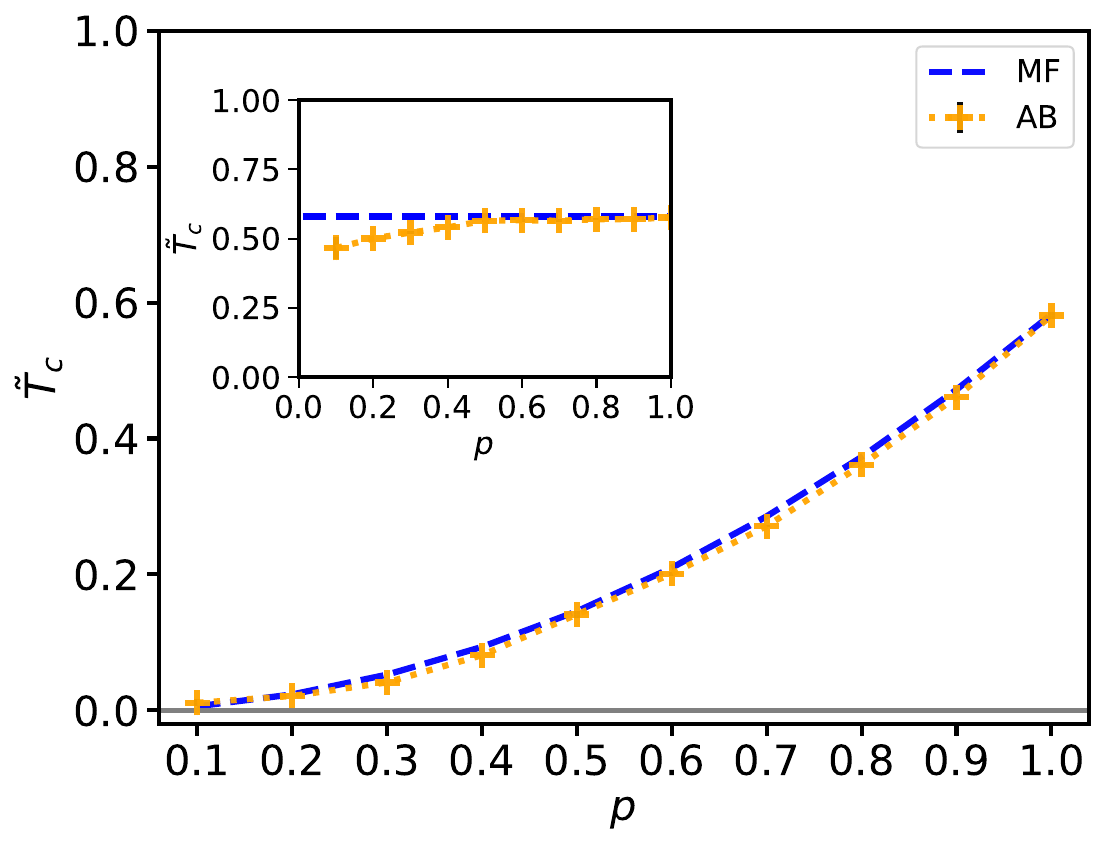}
    \caption{Critical temperature increases as $p^2$ for a single layer ER network started from the paradise state (shown for $N=200$). The dashed blue line represents the analytically predicted normalized critical temperature $\Tilde{T}_c$, and the orange pluses correspond to $\Tilde{T}_c$ values from the simulation. 
    The inset shows the normalized critical temperature observed in simulations where interaction strength is normalized to $A/p^2$. The normalized critical temperature $\Tilde{T}_c$ remains approximately constant for all values of $p$ and has the same value as for the complete graph.
    }
    \label{fig:criticaltemp1L}
\end{figure}

For the paradise state initial condition, the system switches to a disordered state as the temperature surpasses the normalized critical temperature $\Tilde{T}_c$.
Figure \ref{fig:time-evolution} illustrates the temporal evolution of $x$ across different values of normalized temperature $\Tilde{T}$ for a system size of $N=200$. The dashed black line represents the critical value $x_c$ given in sec.\ref{sec:meanfield single}.
Figure \ref{fig:meanfield} shows the mean link polarization and the corresponding normalized intralayer energy $\Tilde{E}^{(1)}$ obtained from Eq.\ref{eq:normalizedE}, for different connection probability $p$ for network sizes $N=200$ and $400$.
The results indicate that the normalized critical temperature $\Tilde{T}_c$ increases as $p$ increases.
The intralayer energy approaches zero, and mean polarization fluctuates around zero above the critical temperature.
The critical temperature obtained from the simulations agrees with the mean-field approximation.\\
A comparison between the normalized critical temperature $\Tilde{T}_c$ predicted by analytical mean-field theory from Eq.\ref{eq:CriticalTemp2L} and those obtained from Monte Carlo simulations is shown in Fig. \ref{fig:criticaltemp1L}.
The system's initial condition is the paradise state, and the transition to a disordered state occurs at the normalized critical temperature $\Tilde{T}_c$, above which the ordered state is unstable.
This observed temperature scales with an exponent $b \approx 2.1$, determined by fitting a power function $\Tilde{T}_c(p)=c p^b$ to the simulation results.
This scaling aligns well with the expected mean-field prediction, which suggests that the critical temperature should scale with $p^2$.
The details of the method used to find critical temperature from the simulation data are discussed in Appendix \ref{sec:appendix}.
Despite the slight discrepancy observed in the transition point when comparing numerical simulations for $p<0.5$ with the mean-field approximation, the analytical method is successful.
The fact that the mean-field approximation becomes less accurate as the network becomes sparser is expected since, by its nature, the mean-field captures the complete graph most accurately.
In fact, the case of $p=0.1$ is special since the sharp first-order transition could not be found in the data, suggesting the disappearance of the expected discontinuous transition, similar to observations of Malarz et al. \cite{malarz2023heiderer}, which means that the agreement between numerical results and analytical predictions should be treated with caution.
The details on this issue are found in Appendix \ref{sec:appendixB}.\\
If the connection probability $p$ is changed and the interaction strength $A$ is simultaneously re-normalized to $A/p^2$, the normalized critical temperature $\tilde{T}_c$ remains constant.
This is shown in the Fig.\ref{fig:criticaltemp1L} inset, where additional simulations were performed with interaction strength $A/p^2$, keeping the observed normalized critical temperature $\tilde{T}_c$ roughly constant, equal to the critical temperature for the complete graph with interaction strength $A$.
Similar to the main graph data, the analytical predictions become less accurate as $p$ decreases.

\subsection{Distribution of triads}
\label{sec:numerical-triads}
\begin{figure}
         \subfloat[\label{fig:subfig6a}]{
     \includegraphics[width=0.47\linewidth]{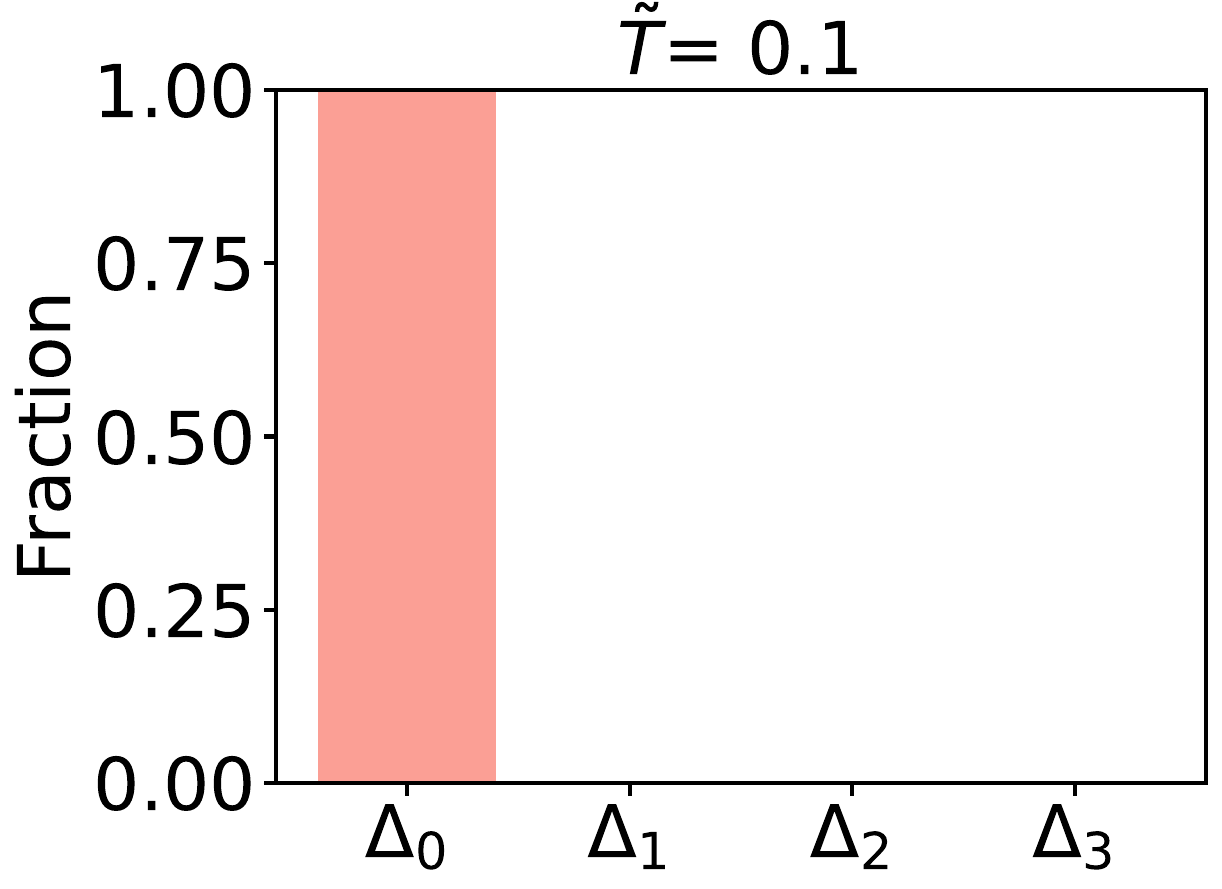}
     }\hfill
    \subfloat[\label{fig:subfig6b}]{
    \includegraphics[width=0.47\linewidth]{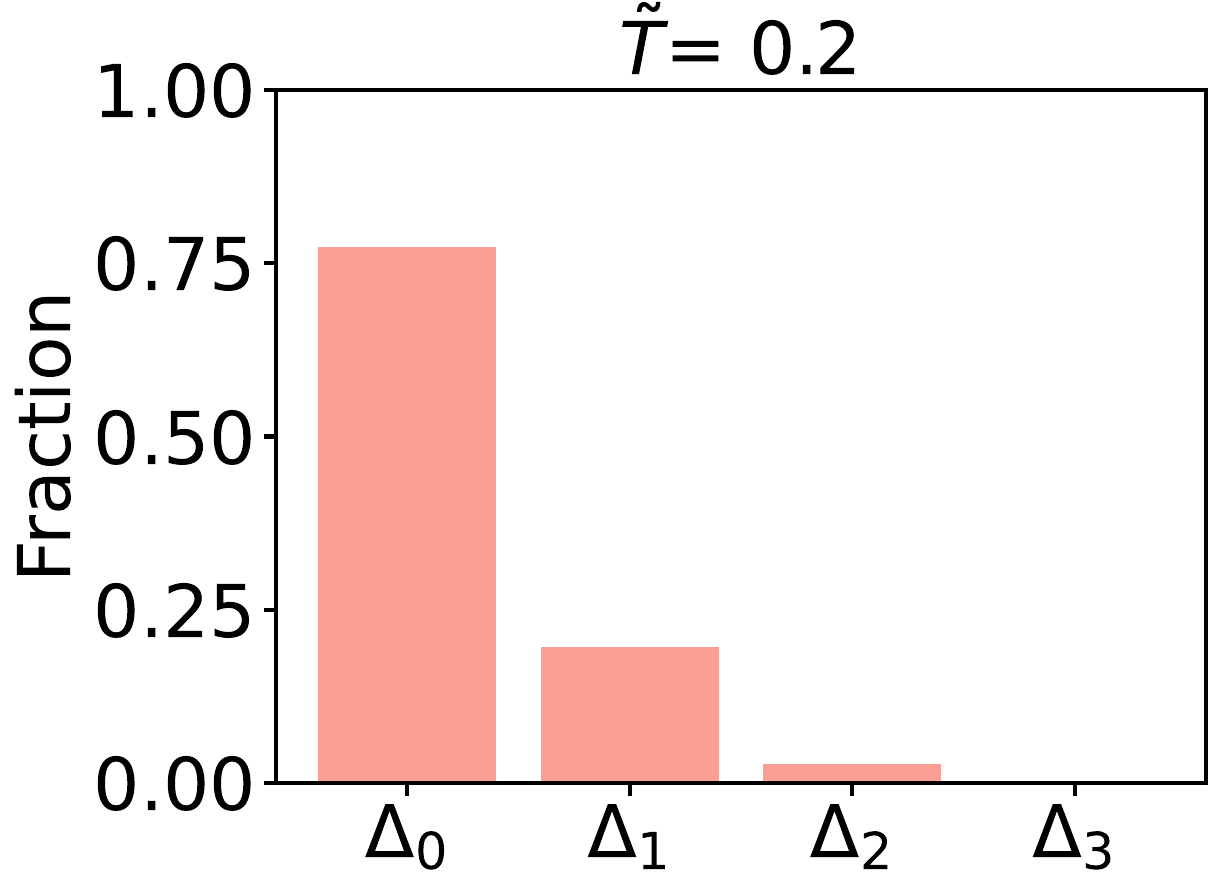}
    }\hfill
    \subfloat[\label{subfig6c}]{\includegraphics[width=0.48\linewidth]{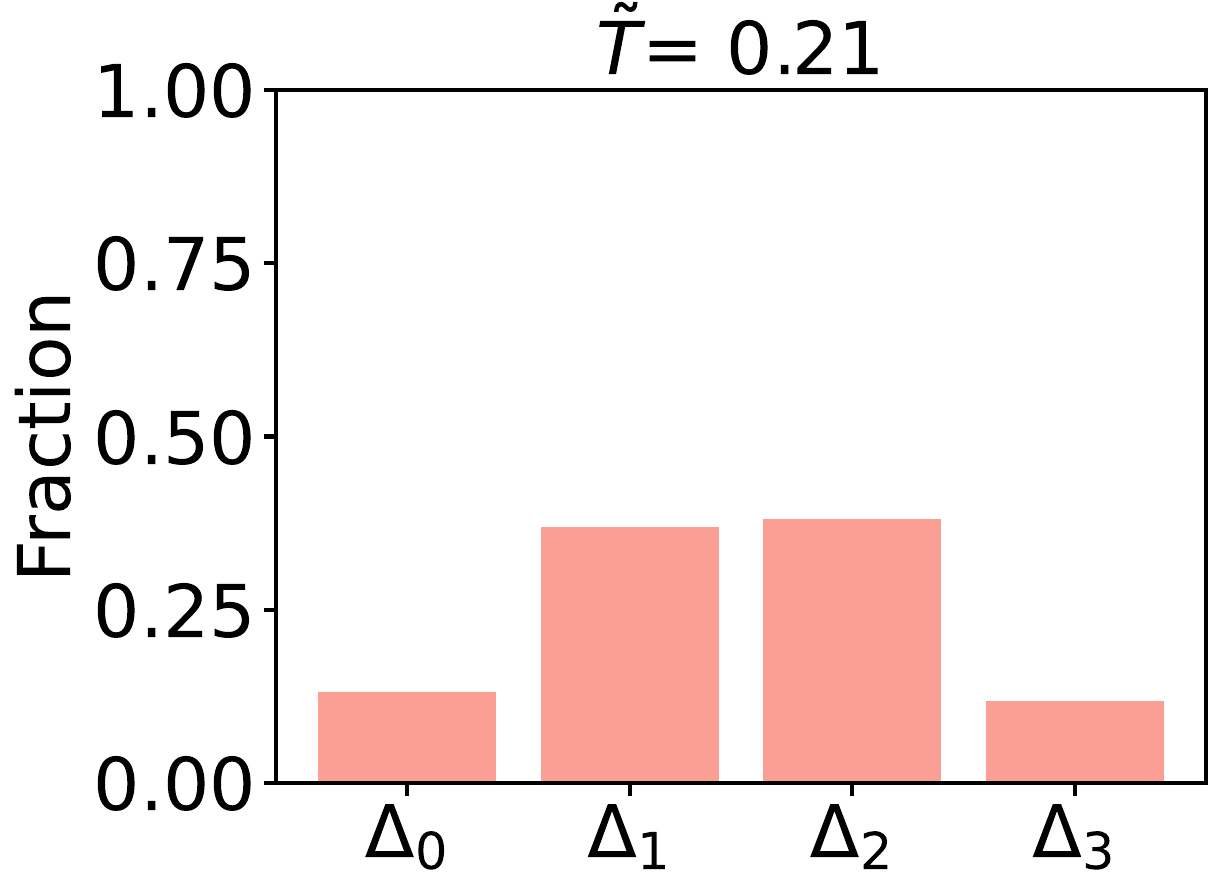}
    }\hfill
    \subfloat[\label{subfig6d}]{\includegraphics[width=0.48\linewidth]{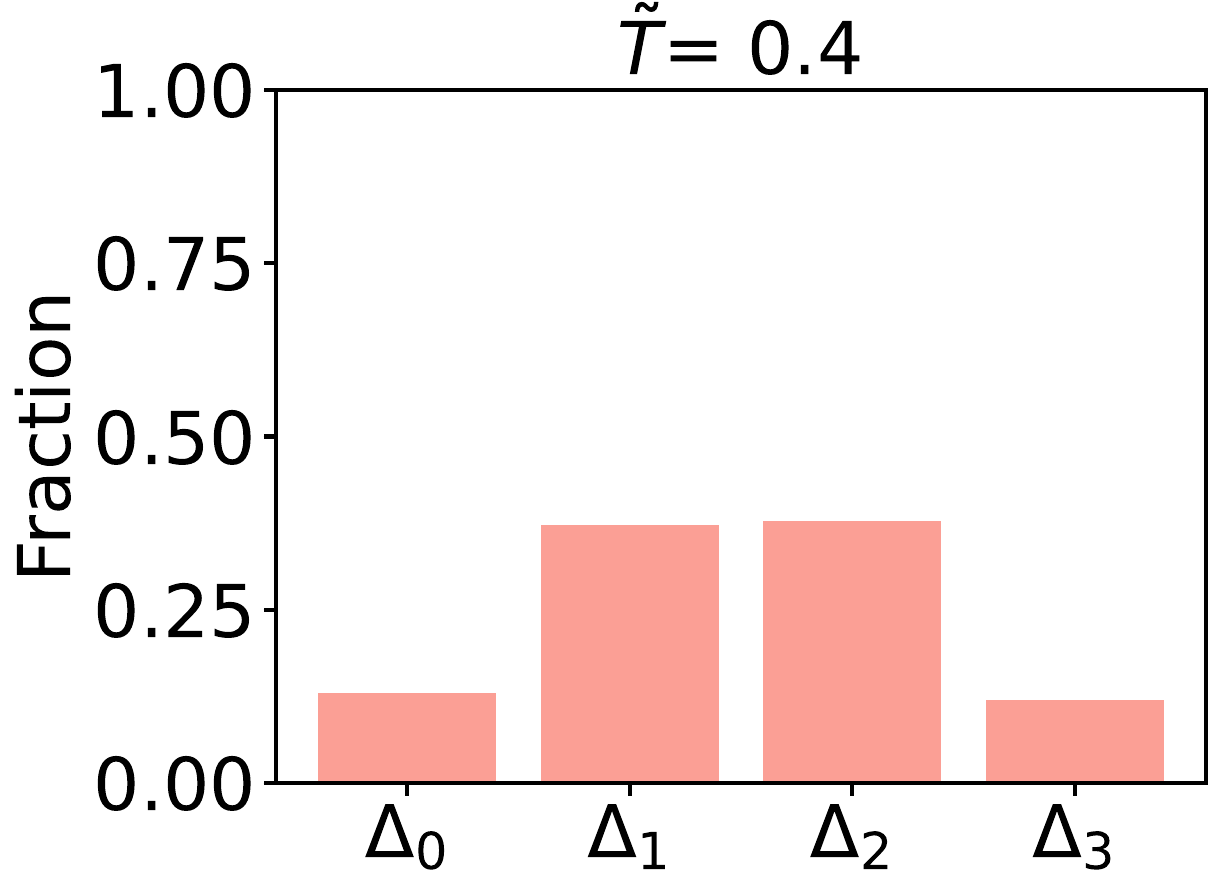}
    }
    \caption{The distribution of triads for the ER graph initialized from the paradise state representing the network's asymptotic state depicted in Fig.\ref{fig:time-evolution} at different temperatures for $N=200$ and $p=0.6$. The ER network started from the paradise state has all its triads in the balanced state $(\Delta_0)$ at low temperatures (a). As the temperature remains below $\Tilde{T}_c\approx 0.2097$, the distribution retains a high proportion of stable $\Delta_0$ triads (b). Above the critical temperature, the distribution shifts dramatically to $\approx \left(\frac{1}{8},\frac{3}{8},\frac{3}{8},\frac{1}{8}\right)$ (c and d).}
    \label{fig:triaddistribution}
\end{figure}

To ensure that our interpretation of the system state based on polarization $x$ and energy $E$ is correct, we investigated the occurrence frequency of different types of triads within the single-layer ER network.
We differentiate triad types by the number of negative links they contain and represent them as $\Delta_{l}$, where the subscript indicates the number of negative links.
The balance of a triad is characterized by the product of the signs of its edges, so a triad of type $\Delta_0$ and $\Delta_2$ (even number of negative links) are balanced, and the triads of type $\Delta_1$ and $\Delta_3$ (odd number of negative links) are unbalanced. 
When the ER graph is initiated with paradise state $(x=1)$, our findings reveal that at low temperatures (as illustrated in Fig. \ref{fig:triaddistribution}), practically all triads are in a balanced $\Delta_0$ configuration, indicating the paradise state.
As long as the temperature remains below the normalized critical temperature $\Tilde{T}_c$, the network maintains a majority of triads of type $\Delta_0$, with fluctuations resulting in a significant fraction of $\Delta_1$.
Once the temperature exceeds $\Tilde{T}_c$, the distribution of triads undergoes a dramatic transformation.
Above the critical temperature, the distribution of triads is dominated by entropy, with triads $\Delta_1$ and $\Delta_2$ being approximately three times more prevalent than $\Delta_0$ and $\Delta_3$, as expected from a random distribution of link signs in a network.
A small influence of Heider balance can still be observed, with balanced triads $\Delta_0$ and $\Delta_2$ being slightly more popular than their imbalanced counterparts $\Delta_3$ and $\Delta_1$.

\subsection{Effect of coupling for a bilayer ER network}
\label{sec:numerical-bilayer}

\begin{figure}
    \subfloat[\label{fig:subfig7a}]{
    \includegraphics[width=0.9\linewidth]{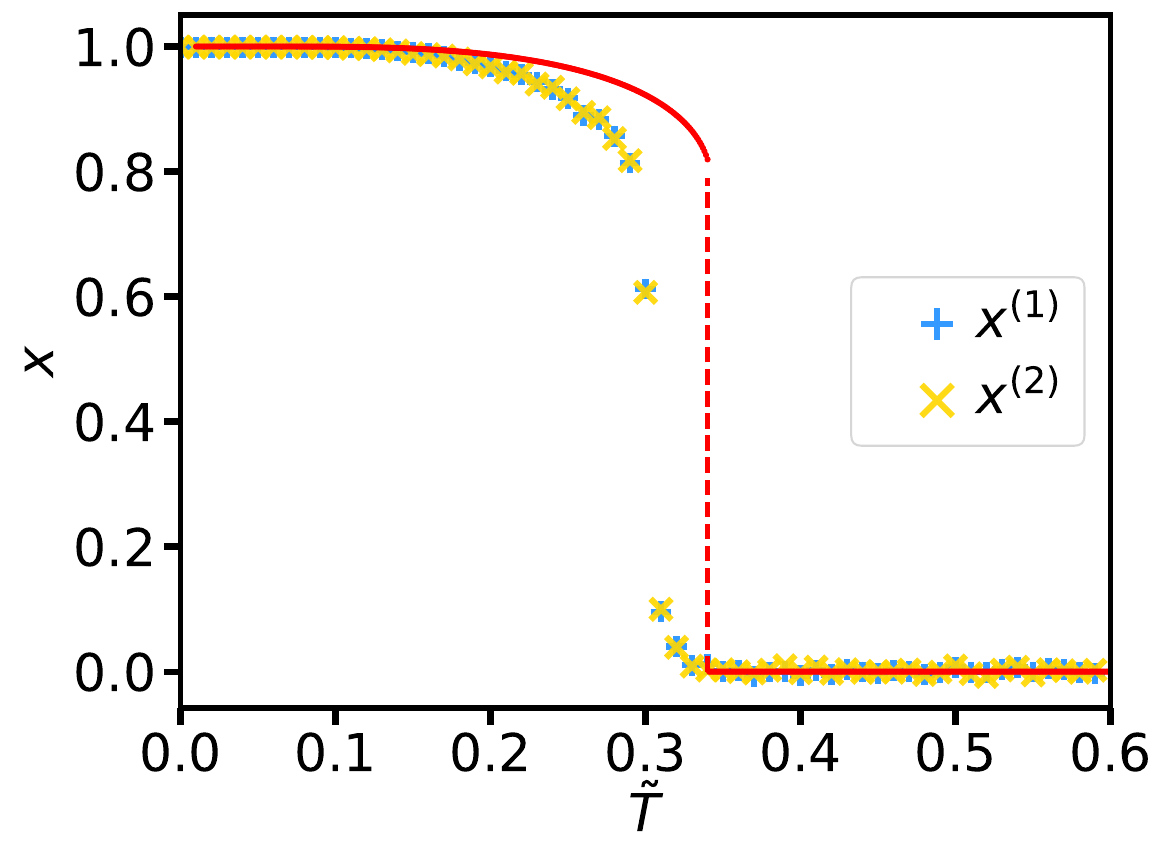}
    }\hfill
    \subfloat[\label{fig:subfig7b}]{
    \includegraphics[width=0.9\linewidth]{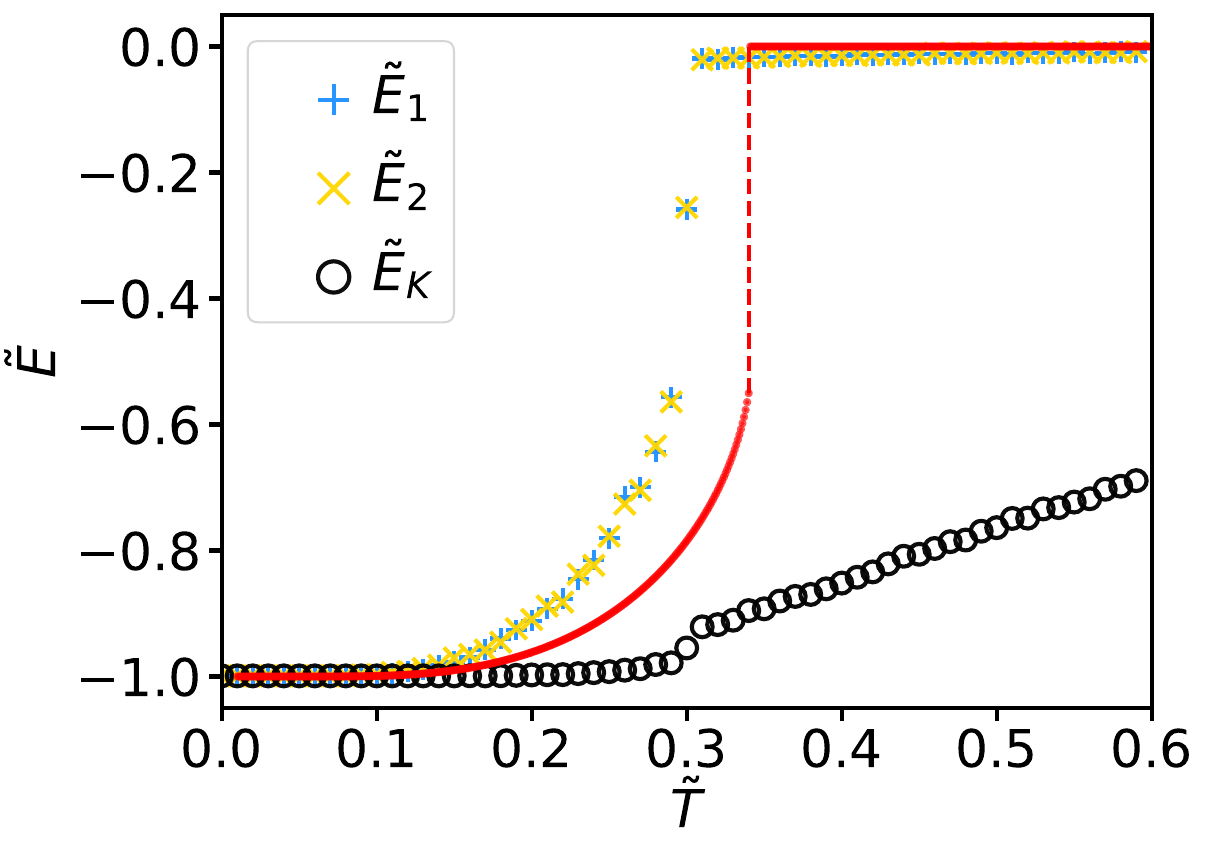}
    }
    \caption{Mean link polarization (a) and corresponding normalized energies (b) for a duplex ER network as a function of the normalized temperature $\tilde{T}=T/(AM)$ for a system of $N = 200$ nodes, with $p=0.6$ and a coupling strength $K = 25$. The results show averages of 50 independent simulations. Initial conditions were paradise states at both layers, where the overlapping blue plus and yellow cross symbols correspond to layers $(1)$ and $(2)$, respectively. The red points at both panels show the analytical results for mean polarization and intralayer energies.
    While normalized intralayer energies $\tilde{E}_1$ and $\tilde{E}_2$ (overlapping blue plus and yellow cross symbols) display a discontinuous transition to around zero, the interlayer energy $E_K$ (black circle) shows only a small change. Then, it increases slowly when $T > \Tilde{T}_c$, indicating that layers remain partially synchronized even in a disordered state.
    }
    \label{fig:bilayer PP}
\end{figure}

\begin{figure}
    \centering
    \includegraphics[width=1\linewidth]{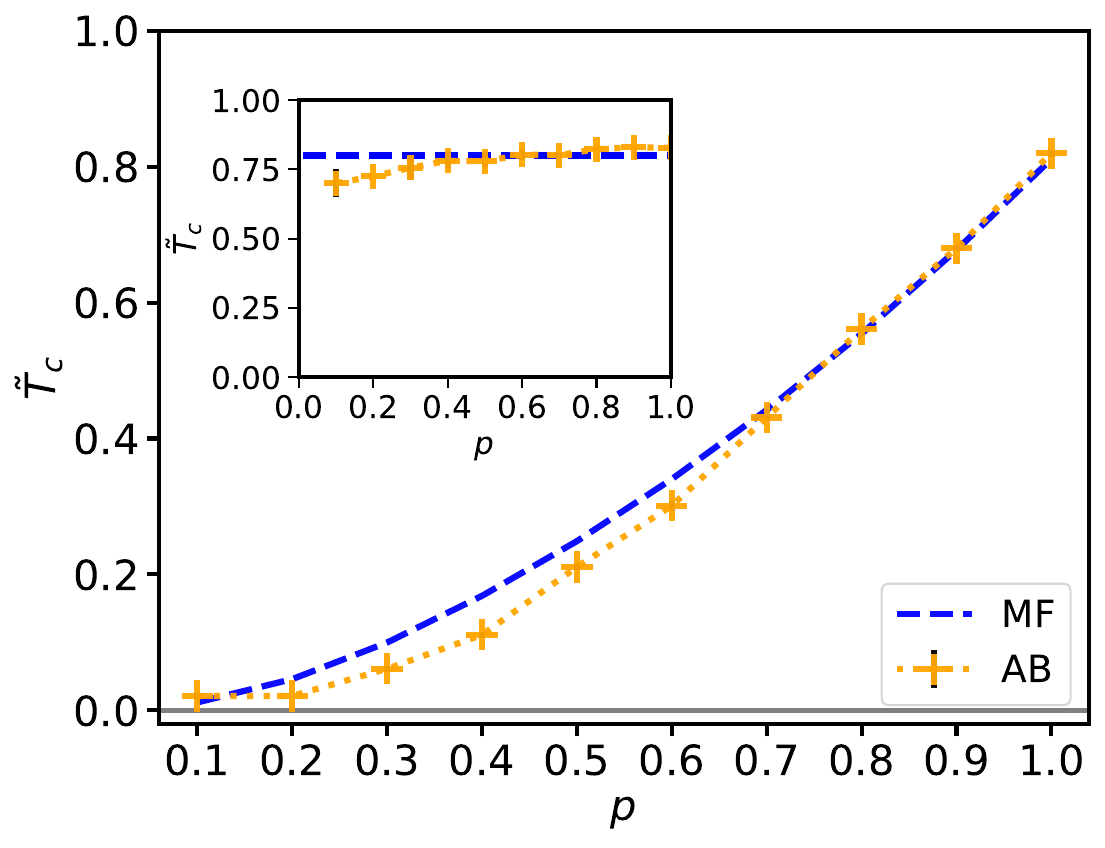}
    \caption{The normalized critical temperature $\tilde{T}_c=T_c/(AM)$ for a bilayer ER network increases with $p$ (shown for $N=200$ and $K=100$). The blue dashed line shows the analytical results, and the orange pluses correspond to the results of agent-based simulation. There is some discrepancy in the critical temperature for $p = 0.1$ and $p = 0.2$.
    The inset shows the situation when intralayer Heider coupling strength is re-scaled to $A/p^2$ and interlayer Ising coupling to $K/p$.
    In such case, the normalized critical temperature $\tilde{T}_c$ remains approximately constant, as predicted analytically.
    }
    \label{fig:criticaltemp2L}
\end{figure}

In the case of a bilayer ER network, if both layers are in the paradise state, then the system is balanced.
This scenario reflects a condition where all intralayer relationships are positive $\langle x_{ij}^{(1)}\rangle=\langle x_{ij}^{(2)}\rangle = 1$, leading to a harmonious and fully cooperative network environment across both layers.
In our simulations, the interaction parameters in both layers of the network are kept as $A^{(1)} = A^{(2)} = 1$, reflecting uniformity in the strength of interactions across the network layers.\\
Up to temperature $\Tilde{T}_c=\Tilde{T}_c(K,p)$, the system remains polarized with $\langle x_{ij}^{(1)}\rangle \approx \langle x_{ij}^{(2)}\rangle > 0$, while above $\Tilde{T}_c$, the system discontinuously changes to a disordered state $\langle x_{ij}^{(1)}\rangle \approx \langle x_{ij}^{(2)}\rangle \approx 0$.
This is shown in Figure \ref{fig:bilayer PP}, where a consistent alignment is observed across both layers of the system below normalized critical temperature $\Tilde{T}_c$, and above $\Tilde{T}_c$, the mean polarization drops to around zero, reflecting a disordered state.
The critical temperature derived from these results is lower than that predicted by the mean-field approximation.
At the same time as a change in polarization occurs, a discontinuous behavior is observed for the normalized intralayer energies $\Tilde{E}_{1}$ and $\Tilde{E}_{2}$, that converge towards zero.
A small local ordering persists, lowering the mean value of energy below zero. 
The normalized interlayer energy $\Tilde{E}_K$ also changes but displays negative values both above and below the critical temperature.
This observation suggests a significant, partial alignment of link signs across different layers, even amidst the disordered state above $\Tilde{T}_c$.
This behavior is similar to the case of a complete graph $(p=1)$ (see ref.\cite{mohandas2024critical})\\
Figure \ref{fig:criticaltemp2L} compares the normalized critical temperature $\Tilde{T}_c=T_c/(AM)$ predicted by the mean-field theory and temperature values obtained from Monte Carlo simulations where both layers started from a paradise state.
The critical temperature increases with $p$ and is in good agreement with the analytical results for high $p$.
The discrepancy for low $p$ appears due to the sparser network behavior not being described well by the mean-field approximation.\\
Suppose the connection probability of the network $p$ is changed, and the intralayer interaction strength is simultaneously re-scaled to $A/p^2$ and the interlayer strength re-normalized to $K/p$. In such a case, the critical temperature estimated from analytical mean-field ~ calculations (Eq. \ref{eq:CriticalTemp2L}) is independent of the parameter $p$, and this prediction is confirmed by numerical simulations presented in the inset of Fig.\ref{fig:criticaltemp2L}.

\section{Conclusions }
We have investigated the behavior of the Heider balance model in Erd\"os-R\'enyi random graphs, in particular its transition from the ordered paradise state to disorder, both in a single-layer and a bilayer network.
We find that the critical temperature of this transition in a single network scales with network density as $p^2$, while for a bilayer network, the scaling becomes much more complex.
In fact, the model with Heider interaction strength $A$ and interlayer interaction strength $K$ behaves the same as in complete graphs with Heider interaction strength $A p^2$ and interlayer interaction strength $K p$, provided that $p$ is not too small. 
This scaling with network density is unlike typical two-body interaction models like the Ising model, where critical temperature simply scales with $p$ linearly \cite{dembo2010ising,dorogovtsev2002ising}.
The difference in re-scaling of intralayer and interlayer interaction strength comes from the fact that the number of triads a given link is involved in scales faster with link density than the probability that a link in one layer will have a matching link in the other layer.
This behavior has been described analytically using a mean-field approach and verified using numerical Monte Carlo simulations, although the prediction is inaccurate for networks with low link density.
The behavior of the Heider structural balance in random graphs has already been investigated before \cite{masoumi2022modified, malarz2023heiderer}.
While \cite{masoumi2022modified} studied the problem, their analytical description focused on critical temperature for transition to disorder in the case of initially disordered networks, which corresponds to the temperature above which disorder becomes a stable state.
Due to system multistability, also described in \cite{masoumi2022modified}, this is not the same as the temperature $\tilde{T}_c$ where order becomes unstable, which we have focused on in our work.
Similarly, \cite{malarz2023heiderer} only investigates the temperature where initial disorder can prevail, not when order disappears, but notes that such a transition becomes continuous if the density of the network $p$ is low enough.
This is similar to the situation observed in our numerical results for $p=0.1$, where the transition from paradise to disorder seems to lose its discontinuous character.\\
The normalized critical temperature $\tilde{T}_c$ we investigated is expected to scale as $p^2$, and a fit to numerical results suggests an even slightly higher exponent. In contrast, the critical temperature found in \cite{malarz2023heiderer} (which we will refer to as $T_d$ to conform with previous literature \cite{malarz2022mean,mohandas2024critical}) shows slower scaling with the exponent $\sim 1.72$.\\
Overall, we have demonstrated that although ER network structure contains a topological disorder, the spontaneous order in structural balance for ER graphs can still exist below a critical temperature, and system critical properties can be analytically described by a properly adapted mean-field approach, similar to the one used for complete graphs.

\begin{acknowledgements}
This research received funding from the Polish National Science Center under Alphorn Grant No. 2019/01/Y/ST2/00058. This work was funded by the European Union under the Horizon Europe grant OMINO (Grant No. 101086321) and was also co-financed with funds from the Polish Ministry of Education and Science under the program entitled International Co-Financed Projects.
\end{acknowledgements}

\appendix
\section{Finding critical temperature from numerical simulations}\label{sec:appendix}

\begin{figure}
    \subfloat[\label{fig:subfig9a}]{
    \includegraphics[width=0.9\linewidth]{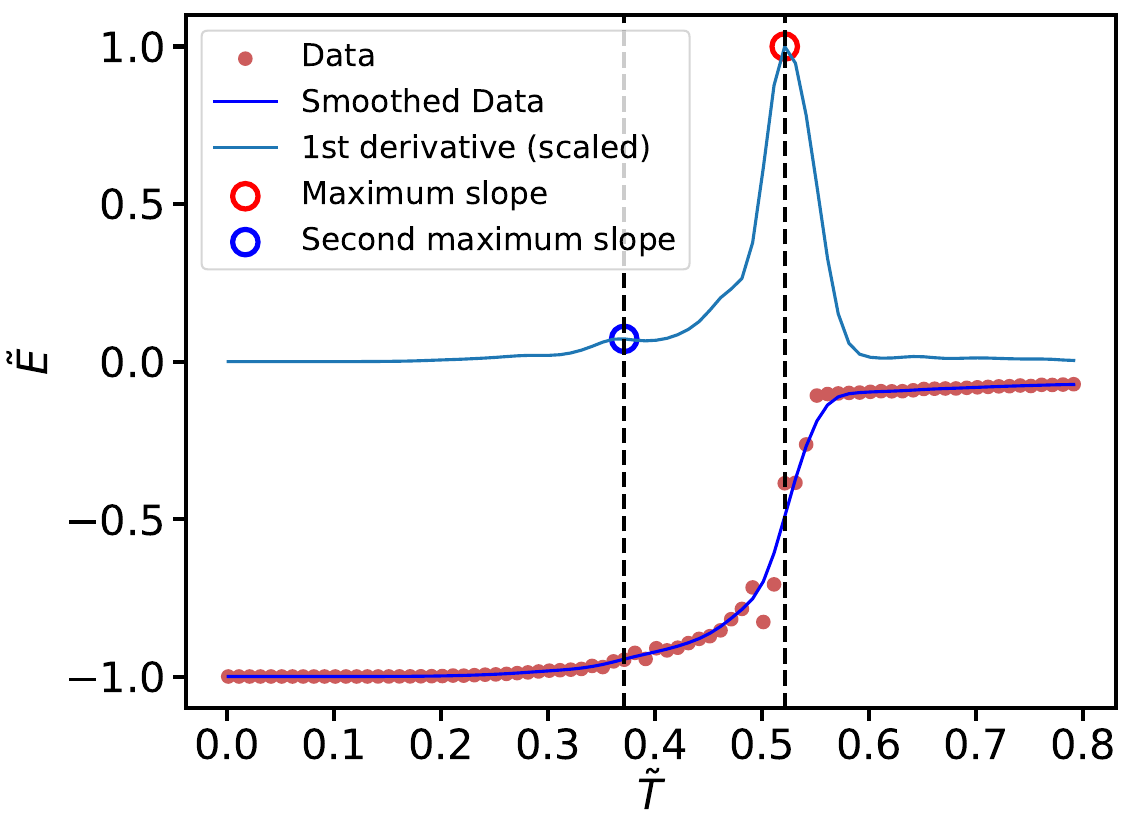}
    }\hfill
    \subfloat[\label{fig:subfig9a}]{
    \includegraphics[width=0.9\linewidth]{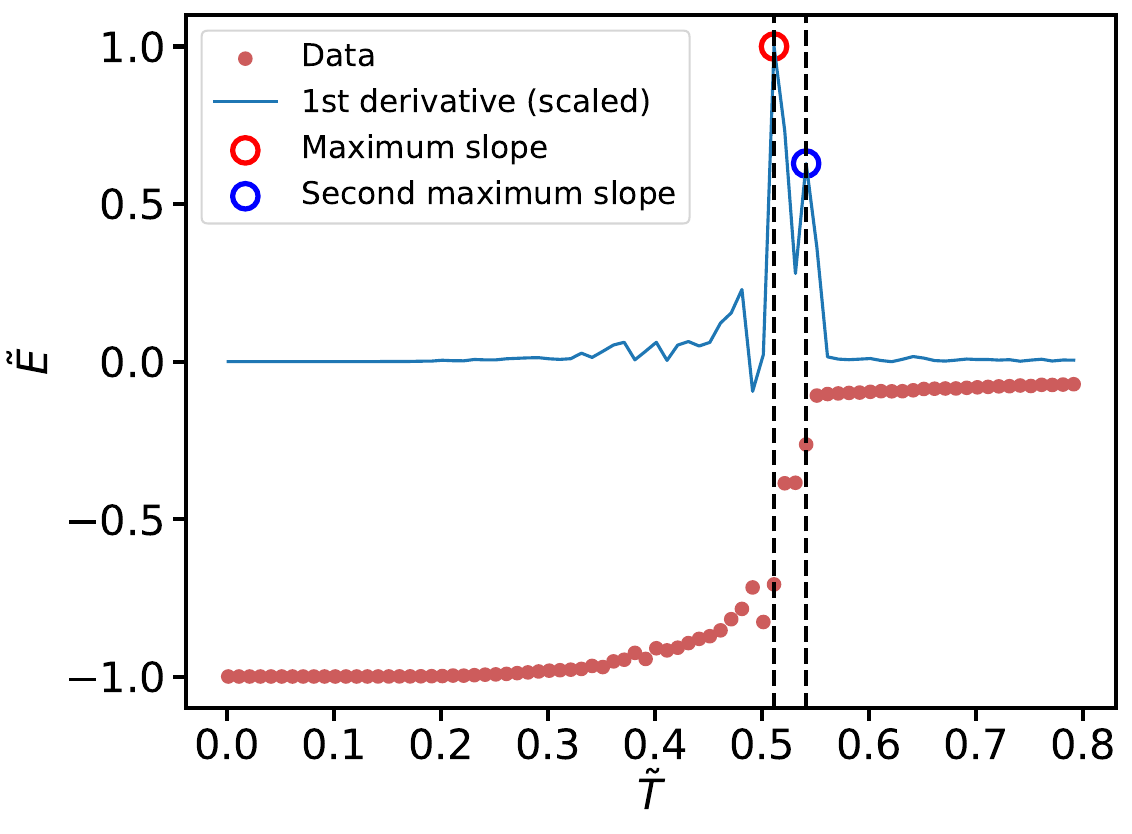}
    }
    \caption{ The critical temperature for a single-layer ER network (normalized interaction strength) of size $N=200$ and $p=0.3$ is identified by examining the maximum of the slope in the dependence of normalized intralayer energy $\Tilde{E}$. This temperature is derived from the maximum of the first derivative, where the ratio of the slopes (red and blue circles) differs between the smoothed data (a) and the unsmoothed data (b). The optimal $\sigma$ for the smoothed data is chosen by examining the ratio of the absolute values of the largest and second-largest first derivatives of smoothed data. In this case, the smoothed data has an optimal  $\sigma$ of  $0.51$.}
    \label{fig:inflection}
\end{figure}

The critical temperature calculated from the simulation is determined by identifying the maximum of the slope where the system's normalized energy, as described by Eq. (\ref{eq:normalizedE}), changes from the ground state to a disordered state with higher energy.
The normalized energy $\Tilde{E}$ is used to identify the critical temperature instead of mean polarization because the energy is less susceptible to fluctuations and early state transition due to finite size effects (partially seen in Fig. \ref{fig:meanfield}), especially for low values of $p$.
This difference in behavior makes energy a more reliable observable for determining the critical temperature.\\
To determine the critical temperature, we apply Gaussian kernel smoothing to the normalized energy data, averaging over $50$ realizations.
A key component in Gaussian smoothing is selecting the appropriate $\sigma$ value, which defines the width of the Gaussian distribution used to smooth the data.
The correct choice of $\sigma$ significantly impacts the clarity and accuracy of the results, especially for low values of $p$ where the transition behavior is rather noisy.\\
To select the optimal value of $\sigma$, we evaluate a range of possible $\sigma$ values starting from $0$.
For each value of $\sigma$, we identify the two largest maxima of the slope.
If there is only one maximum or the ratio between the largest and second-largest maxima is $10$ or more (arbitrary factor), we conclude that the smoothing is optimal, and we have found the one true maximum and take it as our observed critical temperature.
If this condition is not fulfilled, we increase $\sigma$ by $0.1$ and try again until we have smoothed the data enough for the above criteria to be fulfilled.
Figure \ref{fig:inflection} illustrates finding slope maxima for unsmoothed and smoothed data.
Since the original data is discrete, the smoothed data retains this discrete nature.
To find slope maxima, we calculate the numerical derivatives of the smoothed data.
Gradients are computed using central differences for internal points and first differences at the boundaries.

\section{Continuity of transition for sparse graphs}\label{sec:appendixB}
\begin{figure}[H]
    \subfloat[\label{fig:subfig10a}]{
    \includegraphics[width=0.9\linewidth]{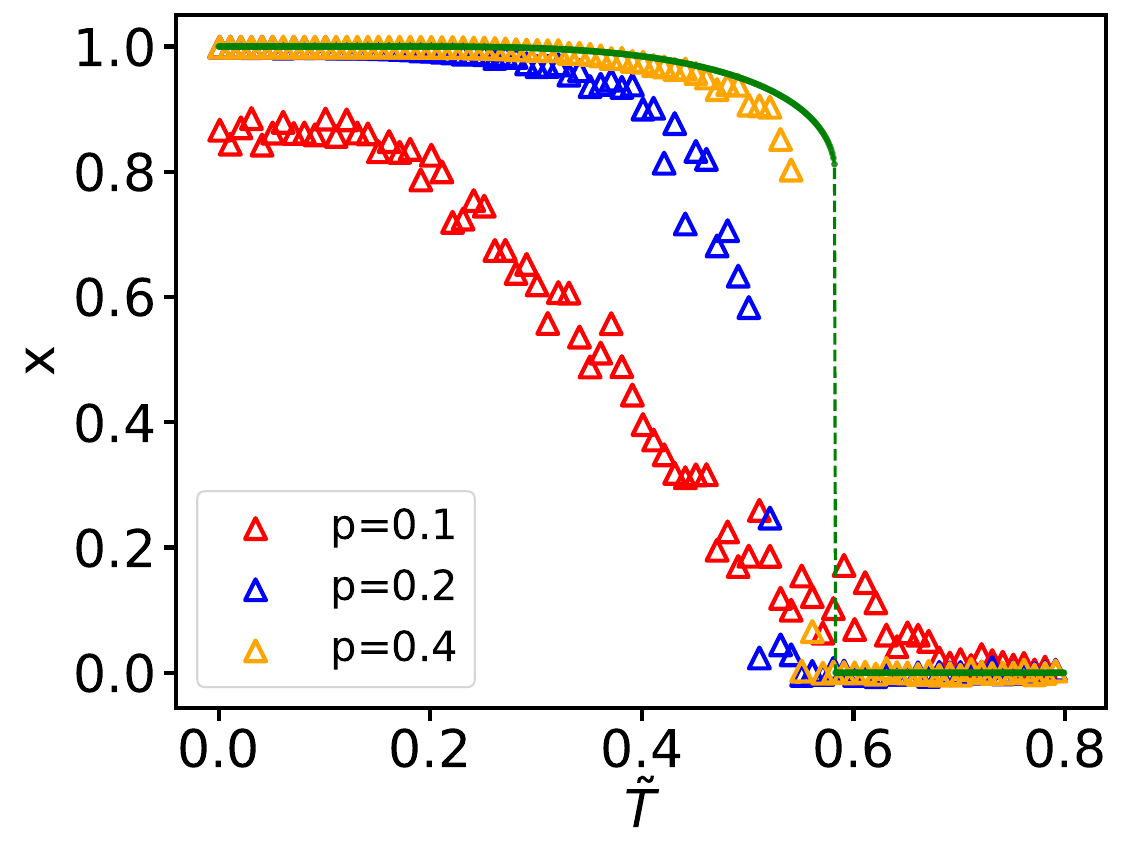}
    }\hfill
    \subfloat[\label{fig:subfig10b}]{
    \includegraphics[width=0.9\linewidth]{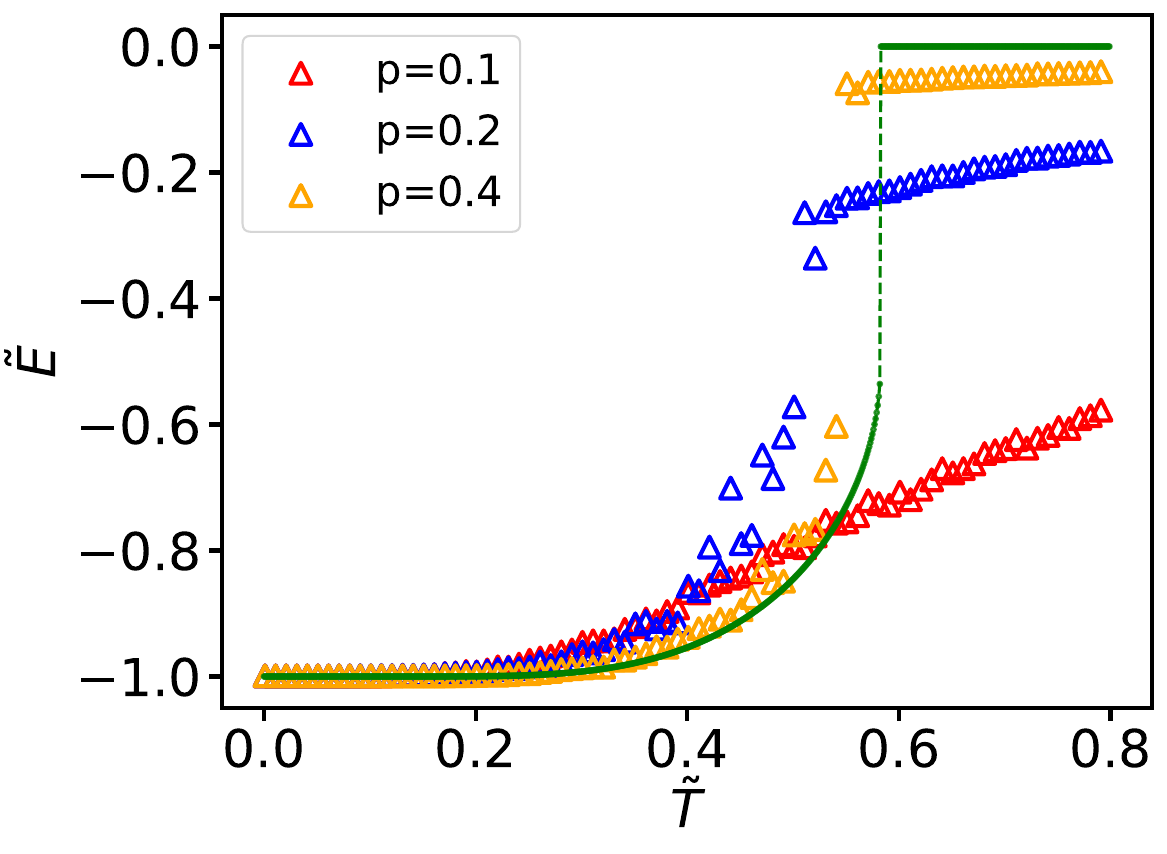}
    }
    \caption{Simulation results for the normalized interaction strength $A/p^2$ for a single-layer ER network of size $N=200$ and different connection probability $p$. A distinct first-order transition is not observed for $p=0.1$ in normalized energy $\Tilde{E}$ and mean polarization $x$ data. Green curves show the  mean-field results that are independent of the parameter  $p$.}
    \label{fig:normalized_X_E}
\end{figure}

In the data obtained from simulations for low $p$, the discontinuity of the transition from ordered to disordered state is not evident.
Due to the close proximity of critical temperatures to zero, this is not very clear for the data underlying the main graph in Fig. \ref{fig:criticaltemp1L}, but it can be seen much clearer for the re-scaled inset data.
Figure \ref{fig:normalized_X_E} shows the behavior of the polarization $x$ and normalized energy $\tilde{E}$ for sparse ER networks when $A/p^2 = \mathrm{const.}$
For $p=0.1$, the transition between ordered and disordered states does not appear to be discontinuous, even if we allow for finite-size effects always present in the simulations.
This means that while we are able to calculate critical temperature $T_c$ for $p=0.1$ using data shown in Fig. \ref{fig:subfig10b}, and their values are relatively close to those expected from the analytical approach, the nature of the transition is not the same and the numerical results should be considered with caution.
Note that although the effect is similar to that observed by Malarz et al. \cite{malarz2023heiderer} for a disorder to order transition, the transition investigated is different, which means neither our results can be considered a confirmation of that of Malarz et al. nor vice versa.

\bibliography{main}
\end{document}